
%
%
\def\unredoffs{} \def\redoffs{\voffset=-.31truein\hoffset=-.48truein}
\def\speclscape{}
%
%
%
%
%
\newbox\leftpage \newdimen\fullhsize \newdimen\hstitle \newdimen\hsbody
\tolerance=1000\hfuzz=2pt
\catcode`\@=11 
\def\bigans{b }
\def\answ{b }
%
\ifx\answ\bigans\message{(This will come out unreduced.}
\magnification=1200\unredoffs\baselineskip=16pt plus 2pt minus 1pt
\hsbody=\hsize \hstitle=\hsize 
\else\message{(This will be reduced.} \let\l@r=L
\magnification=1000\baselineskip=16pt plus 2pt minus 1pt \vsize=7truein
\redoffs \hstitle=8truein\hsbody=4.75truein\fullhsize=10truein\hsize=\hsbody
\output={\ifnum\pageno=0 
  \shipout\vbox{\speclscape{\hsize\fullhsize\makeheadline}
    \hbox to \fullhsize{\hfill\pagebody\hfill}}\advancepageno
  \else
  \almostshipout{\leftline{\vbox{\pagebody\makefootline}}}\advancepageno
  \fi}
\def\almostshipout#1{\if L\l@r \count1=1 \message{[\the\count0.\the\count1]}
      \global\setbox\leftpage=#1 \global\let\l@r=R
 \else \count1=2
  \shipout\vbox{\speclscape{\hsize\fullhsize\makeheadline}
      \hbox to\fullhsize{\box\leftpage\hfil#1}}  \global\let\l@r=L\fi}
\fi
%
\newcount\yearltd\yearltd=\year\advance\yearltd by -1900

\def\Title#1#2{\nopagenumbers\abstractfont\hsize=\hstitle\rightline{#1}%
\vskip 1in\centerline{\titlefont #2}\abstractfont\vskip .5in\pageno=0}
\def\Date#1{\vfill\leftline{#1}\tenpoint\supereject\global\hsize=\hsbody%
\footline={\hss\tenrm\folio\hss}}
%

\def\draftmode{\message{ DRAFTMODE }\def\draftdate{{\rm preliminary draft:
\number\month/\number\day/\number\yearltd\ \ \hourmin}}%
\headline={\hfil\draftdate}\writelabels\baselineskip=20pt plus 2pt minus 2pt
 {\count255=\time\divide\count255 by 60 \xdef\hourmin{\number\count255}
  \multiply\count255 by-60\advance\count255 by\time
  \xdef\hourmin{\hourmin:\ifnum\count255<10 0\fi\the\count255}}}
\def\nolabels{\def\wrlabeL##1{}\def\eqlabeL##1{}\def\reflabeL##1{}}
\def\writelabels{\def\wrlabeL##1{\leavevmode\vadjust{\rlap{\smash%
{\line{{\escapechar=` \hfill\rlap{\sevenrm\hskip.03in\string##1}}}}}}}%
\def\eqlabeL##1{{\escapechar-1\rlap{\sevenrm\hskip.05in\string##1}}}%
\def\reflabeL##1{\noexpand\llap{\noexpand\sevenrm\string\string\string##1}}}
\nolabels
%
\global\newcount\secno \global\secno=0
\global\newcount\meqno \global\meqno=1
\def\newsec#1{\global\advance\secno by1\message{(\the\secno. #1)}
\global\subsecno=0\eqnres@t\noindent{\bf\the\secno. #1}
\writetoca{{\secsym} {#1}}\par\nobreak\medskip\nobreak}
\def\eqnres@t{\xdef\secsym{\the\secno.}\global\meqno=1\bigbreak\bigskip}
\def\sequentialequations{\def\eqnres@t{\bigbreak}}\xdef\secsym{}
\global\newcount\subsecno \global\subsecno=0
\def\subsec#1{\global\advance\subsecno by1\message{(\secsym\the\subsecno. #1)}
\ifnum\lastpenalty>9000\else\bigbreak\fi
\noindent{\it\secsym\the\subsecno. #1}\writetoca{\string\quad
{\secsym\the\subsecno.} {#1}}\par\nobreak\medskip\nobreak}
\def\appendix#1#2{\global\meqno=1\global\subsecno=0\xdef\secsym{\hbox{#1.}}
\bigbreak\bigskip\noindent{\bf Appendix #1. #2}\message{(#1. #2)}
\writetoca{Appendix {#1.} {#2}}\par\nobreak\medskip\nobreak}
%
%
\def\eqnn#1{\xdef #1{(\secsym\the\meqno)}\writedef{#1\leftbracket#1}%
\global\advance\meqno by1\wrlabeL#1}
\def\eqna#1{\xdef #1##1{\hbox{$(\secsym\the\meqno##1)$}}
\writedef{#1\numbersign1\leftbracket#1{\numbersign1}}%
\global\advance\meqno by1\wrlabeL{#1$\{\}$}}
\def\eqn#1#2{\xdef #1{(\secsym\the\meqno)}\writedef{#1\leftbracket#1}%
\global\advance\meqno by1$$#2\eqno#1\eqlabeL#1$$}
%
\newskip\footskip\footskip14pt plus 1pt minus 1pt 
\def\footnotefont{\ninepoint}\def\f@t#1{\footnotefont #1\@foot}
\def\f@@t{\baselineskip\footskip\bgroup\footnotefont\aftergroup\@foot\let\next}
\setbox\strutbox=\hbox{\vrule height9.5pt depth4.5pt width0pt}
\global\newcount\ftno \global\ftno=0
\def\foot{\global\advance\ftno by1\footnote{$^{\the\ftno}$}}
%
\newwrite\ftfile
\def\footend{\def\foot{\global\advance\ftno by1\chardef\wfile=\ftfile
$^{\the\ftno}$\ifnum\ftno=1\immediate\openout\ftfile=foots.tmp\fi%
\immediate\write\ftfile{\noexpand\smallskip%
\noexpand\item{f\the\ftno:\ }\pctsign}\findarg}%
\def\footatend{\vfill\eject\immediate\closeout\ftfile{\parindent=20pt
\centerline{\bf Footnotes}\nobreak\bigskip\input foots.tmp }}}
\def\footatend{}
%
%
\global\newcount\refno \global\refno=1
\newwrite\rfile
\def\ref{[\the\refno]\nref}
\def\nref#1{\xdef#1{[\the\refno]}\writedef{#1\leftbracket#1}%
\ifnum\refno=1\immediate\openout\rfile=refs.tmp\fi
\global\advance\refno by1\chardef\wfile=\rfile\immediate
\write\rfile{\noexpand\item{#1\ }\reflabeL{#1\hskip.31in}\pctsign}\findarg}
\def\findarg#1#{\begingroup\obeylines\newlinechar=`\^^M\pass@rg}
{\obeylines\gdef\pass@rg#1{\writ@line\relax #1^^M\hbox{}^^M}%
\gdef\writ@line#1^^M{\expandafter\toks0\expandafter{\striprel@x #1}%
\edef\next{\the\toks0}\ifx\next\em@rk\let\next=\endgroup\else\ifx\next\empty%
\else\immediate\write\wfile{\the\toks0}\fi\let\next=\writ@line\fi\next\relax}}
\def\striprel@x#1{} \def\em@rk{\hbox{}}
\def\lref{\begingroup\obeylines\lr@f}
\def\lr@f#1#2{\gdef#1{\ref#1{#2}}\endgroup\unskip}

\def\addref#1{\immediate\write\rfile{\noexpand\item{}#1}} 
\def\startrefs#1{\immediate\openout\rfile=refs.tmp\refno=#1}
\def\xref{\expandafter\xr@f}\def\xr@f[#1]{#1}
\def\refs#1{\count255=1[\r@fs #1{\hbox{}}]}
\def\r@fs#1{\ifx\und@fined#1\message{reflabel \string#1 is undefined.}%
\nref#1{need to supply reference \string#1.}\fi%
\vphantom{\hphantom{#1}}\edef\next{#1}\ifx\next\em@rk\def\next{}%
\else\ifx\next#1\ifodd\count255\relax\xref#1\count255=0\fi%
\else#1\count255=1\fi\let\next=\r@fs\fi\next}
%

%
\newwrite\ffile\global\newcount\figno \global\figno=1
\def\fig{fig.~\the\figno\nfig}
\def\nfig#1{\xdef#1{fig.~\the\figno}%
\writedef{#1\leftbracket fig.\noexpand~\the\figno}%
\ifnum\figno=1\immediate\openout\ffile=figs.tmp\fi\chardef\wfile=\ffile%
\immediate\write\ffile{\noexpand\medskip\noexpand\item{Fig.\ \the\figno. }
\reflabeL{#1\hskip.55in}\pctsign}\global\advance\figno by1\findarg}
\def\xfig{\expandafter\xf@g}\def\xf@g fig.\penalty\@M\ {}
\def\figs#1{figs.~\f@gs #1{\hbox{}}}
\def\f@gs#1{\edef\next{#1}\ifx\next\em@rk\def\next{}\else
\ifx\next#1\xfig #1\else#1\fi\let\next=\f@gs\fi\next}
\newwrite\lfile
{\escapechar-1\xdef\pctsign{\string\%}\xdef\leftbracket{\string\{}
\xdef\rightbracket{\string\}}\xdef\numbersign{\string\#}}

\def\writestop{\def\writestoppt{\immediate\write\lfile{\string\pageno%
\the\pageno\string\startrefs\leftbracket\the\refno\rightbracket%
\string\def\string\secsym\leftbracket\secsym\rightbracket%
\string\secno\the\secno\string\meqno\the\meqno}\immediate\closeout\lfile}}
\def\writestoppt{}\def\writedef#1{}
\def\seclab#1{\xdef #1{\the\secno}\writedef{#1\leftbracket#1}\wrlabeL{#1=#1}}
\def\subseclab#1{\xdef #1{\secsym\the\subsecno}%
\writedef{#1\leftbracket#1}\wrlabeL{#1=#1}}
\newwrite\tfile \def\writetoca#1{}
\def\leaderfill{\leaders\hbox to 1em{\hss.\hss}\hfill}
\def\writetoc{\immediate\openout\tfile=toc.tmp
   \def\writetoca##1{{\edef\next{\write\tfile{\noindent ##1
   \string\leaderfill {\noexpand\number\pageno} \par}}\next}}}

%
\catcode`\@=12 
%
\edef\tfontsize{\ifx\answ\bigans scaled\magstep3\else scaled\magstep4\fi}
\font\titlerm=cmr10 \tfontsize \font\titlerms=cmr7 \tfontsize
\font\titlermss=cmr5 \tfontsize \font\titlei=cmmi10 \tfontsize
\font\titleis=cmmi7 \tfontsize \font\titleiss=cmmi5 \tfontsize
\font\titlesy=cmsy10 \tfontsize \font\titlesys=cmsy7 \tfontsize
\font\titlesyss=cmsy5 \tfontsize \font\titleit=cmti10 \tfontsize
\skewchar\titlei='177 \skewchar\titleis='177 \skewchar\titleiss='177
\skewchar\titlesy='60 \skewchar\titlesys='60 \skewchar\titlesyss='60
\def\titlefont{\def\rm{\fam0\titlerm}
\textfont0=\titlerm \scriptfont0=\titlerms \scriptscriptfont0=\titlermss
\textfont1=\titlei \scriptfont1=\titleis \scriptscriptfont1=\titleiss
\textfont2=\titlesy \scriptfont2=\titlesys \scriptscriptfont2=\titlesyss
\textfont\itfam=\titleit \def\it{\fam\itfam\titleit}\rm}
 \ifx\answ\bigans\else scaled\magstep1\fi
\ifx\answ\bigans\def\abstractfont{\tenpoint}\else
\font\abssl=cmsl10 scaled \magstep1
\font\absrm=cmr10 scaled\magstep1 \font\absrms=cmr7 scaled\magstep1
\font\absrmss=cmr5 scaled\magstep1 \font\absi=cmmi10 scaled\magstep1
\font\absis=cmmi7 scaled\magstep1 \font\absiss=cmmi5 scaled\magstep1
\font\abssy=cmsy10 scaled\magstep1 \font\abssys=cmsy7 scaled\magstep1
\font\abssyss=cmsy5 scaled\magstep1 \font\absbf=cmbx10 scaled\magstep1
\skewchar\absi='177 \skewchar\absis='177 \skewchar\absiss='177
\skewchar\abssy='60 \skewchar\abssys='60 \skewchar\abssyss='60
\def\abstractfont{\def\rm{\fam0\absrm}
\textfont0=\absrm \scriptfont0=\absrms \scriptscriptfont0=\absrmss
\textfont1=\absi \scriptfont1=\absis \scriptscriptfont1=\absiss
\textfont2=\abssy \scriptfont2=\abssys \scriptscriptfont2=\abssyss
\textfont\itfam=\bigit \def\it{\fam\itfam\bigit}\def\footnotefont{\tenpoint}%
\textfont\slfam=\abssl \def\sl{\fam\slfam\abssl}%
\textfont\bffam=\absbf \def\bf{\fam\bffam\absbf}\rm}\fi
\def\tenpoint{\def\rm{\fam0\tenrm}
\textfont0=\tenrm \scriptfont0=\sevenrm \scriptscriptfont0=\fiverm
\textfont1=\teni  \scriptfont1=\seveni  \scriptscriptfont1=\fivei
\textfont2=\tensy \scriptfont2=\sevensy \scriptscriptfont2=\fivesy
\textfont\itfam=\tenit \def\it{\fam\itfam\tenit}\def\footnotefont{\ninepoint}%
\textfont\bffam=\tenbf \def\bf{\fam\bffam\tenbf}\def\sl{\fam\slfam\tensl}\rm}
\font\ninerm=cmr9 \font\sixrm=cmr6 \font\ninei=cmmi9 \font\sixi=cmmi6
\font\ninesy=cmsy9 \font\sixsy=cmsy6 \font\ninebf=cmbx9
\font\nineit=cmti9 \font\ninesl=cmsl9 \skewchar\ninei='177
\skewchar\sixi='177 \skewchar\ninesy='60 \skewchar\sixsy='60
\def\ninepoint{\def\rm{\fam0\ninerm}
\textfont0=\ninerm \scriptfont0=\sixrm \scriptscriptfont0=\fiverm
\textfont1=\ninei \scriptfont1=\sixi \scriptscriptfont1=\fivei
\textfont2=\ninesy \scriptfont2=\sixsy \scriptscriptfont2=\fivesy
\textfont\itfam=\ninei \def\it{\fam\itfam\nineit}\def\sl{\fam\slfam\ninesl}%
\textfont\bffam=\ninebf \def\bf{\fam\bffam\ninebf}\rm}
%
%

\hyphenation{anom-aly anom-alies coun-ter-term coun-ter-terms}
\def\inv{^{\raise.15ex\hbox{${\scriptscriptstyle -}$}\kern-.05em 1}}

\def\Dsl{\,\raise.15ex\hbox{/}\mkern-13.5mu D} 
\def\dsl{\raise.15ex\hbox{/}\kern-.57em\partial}
\def\del{\partial}

\font\bigit=cmti10 scaled \magstep1
\def\lspace{\ifx\answ\bigans{}\else\qquad\fi}
\def\lbspace{\ifx\answ\bigans{}\else\hskip-.2in\fi} 
\def\boxeqn#1{\vcenter{\vbox{\hrule\hbox{\vrule\kern3pt\vbox{\kern3pt
	\hbox{${\displaystyle #1}$}\kern3pt}\kern3pt\vrule}\hrule}}}
\def\mbox#1#2{\vcenter{\hrule \hbox{\vrule height#2in
		\kern#1in \vrule} \hrule}}  
%

\def\darr#1{\raise1.5ex\hbox{$\leftrightarrow$}\mkern-16.5mu #1}

\def\roughly#1{\raise.3ex\hbox{$#1$\kern-.75em\lower1ex\hbox{$\sim$}}}
\Title{HUTP-92/A055}{On the spectra of
$\widehat{sl}(N)_k/\widehat{sl}(N)_k$-cosets and $W_N$ gravities}

\centerline{Vladimir Sadov} \bigskip\centerline{Lyman Laboratory of
Physics} \centerline{Harvard University}\centerline{Cambridge, MA 02138}
\centerline{and}\centerline{L.~D.~Landau Institute for Theoretical Physics,
Moscow} 
\vskip .3in

We study the spectra of G/G coset models by computing BRST cohomology of
affine Lie algebras with coefficients in tensor product of two modules.
One-to-one correspondence between the spectra of $A_1^1/A_1^1$ and that of
the  minimal matter coupled to gravity (including boundary states of the
Kac table)  is observed. This phenomena is discussed from the point of
hamiltonian reduction  of BRST complexes of $A_N^1$ Lie algebras.

\Date{10/92}

\vfill\eject

\newsec{Introduction}

It has been clear for some time that there should be some relationship
between  noncritical $W_N$ string and
$\widehat{sl}(N)\widehat{sl}(N)_k/\widehat{sl}(N)_k$-cosets. For the
definitions  of both (and for some exposition) we refer reader to [1-4,10].
It
is well known that the hamiltonian reduction [5,6] maps the
representations of  $\widehat{sl}(N)_k$ to that of $W_N$. Loosely speaking,
this procedure kills  the degrees of freedom, corresponding to currents
from nilpotent subalgebra  $N_+$, leaving gauge symmetry enough to fix all
other but $N-1$ currents.  The survivors form $W_N$ algebra.

Naively, considering both $G/G$ coset or some $W_N$ matter coupled to gravity,
we are trying to kill {\it all} degrees of freedom. Then, it seems very
natural to
suppose that there is no difference: either we do reduction first and then
compute
$W_N$-BRST homology to kill the survivors after reduction, or just do
$\widehat{sl}(N)$-
BRST homology to kill all at once.

This naive conclusion, surprisingly, has found some support from comparing
the  results for BRST homologies of Virasoro algebra on one hand [7], and
$\widehat{sl}(2)$  algebra on the other [1,3]. To specify the statement,
let's consider two  modules, $M_1$  and $M_2$ of  $\widehat{sl}(N)$
($\widehat{sl}(2)$ in this example), and two modules  $M_1^R$ and $M_2^R$
of $W_N$ corresponding  to the first pair by hamiltonian reduction.
Usually, $M_1$ is an admissible representation ($k_1+N={p\over q}$ ---
rational ),  and $M_2$ is a Wakimoto ( free fields ) representation with
the value of the  central charge $k_2=-k_1-2N$.

Then, the $\widehat{sl}(2)$-BRST homology of $M_1 \otimes M_2$ does
coincide with  $W_2={\rm Vir}$-BRST homology of  $M_1^R \otimes M_2^R$,
which can be shown by the  direct  computation of both.

In section 2 of this work we will compute the spectra of
$\widehat{sl}(N)_k/\widehat{sl}(N)_k$-coset models
(=BRST homology of $M_1 \otimes M_2$) for several choices for $M_1, M_2$.
(We would like to mention here, that we include in the definition all
homologies, not only (co-)invariants $H_0$. So defined, the spectra turn
out to be infinite in the most interesting cases.)

We will observe, that the homologies can be relatively easily found for the
modules,  corresponding to the  main grid of the Kac table (=operators of
minimal models) (see also [2,3])  Some more elaboration is required to find
the homologies for those,  corresponding to the  ``boundary" of the Kac
table. For Virasoro algebra, these are essentially the   ``discrete states"
of $c=1$ noncritical string. Up to now, these objects have been missed in
analysis even of $\widehat{sl}(2)_k/\widehat{sl}(2)_k$-cosets. We wish  to
stress  the importance of the fact that these states present in cosets,
because we know  that they exist in $W_N$ gravity and that this gives an
additional piece of evidence  in favour of naive assumption, mentioned
above.

Whereas everything is beautiful with the spectra of cosets, it is quite the
opposite  with the spectra of $W_N$-gravity. Although it is really possible
to construct  a  BRST complex in that case [10], it is not at all clear why
it is possible.  Then, it  appears that this complex is not very convenient
for the direct computations.  Nevertheless, for the simplest case of $W_3$
and when there are no ``discrete  states",  it can be shown that the
homology of such complex coincide with that of  $\widehat{sl}(3)$ with the
correspondence of modules described above.

Taken together, all these facts motivate an attempt to define a procedure
of  hamiltonian reduction not only of $\widehat{sl}(N)$ single module, but
also of the  whole BRST complex. For the BRST complex for the product $M_1
\otimes M_2$, it  cannot  be just independent reduction of each module, the
real thing should involve  ghosts  and should be compatible with the
structure of the complex. In Section 3 we address this  issue and give a proper
modification of the reduction procedure for $A_1^1$.

\newsec{Spectrum of $A^1_N/A^1_N$}
\subsec{Some definitions and notations}

We define the Kac-Moody Lie algebra $A^1_N$ by the relations:

$$ [e^\alpha
_n,f^\beta _m]=\delta ^{\alpha \beta}h^\alpha _{n+m} +   kn\delta ^{\alpha
\beta}\delta _{m+n,0}$$  $$[h^\alpha _n, e^\beta _m]=C^{\alpha
\beta}e^\beta _{m+n}$$  $$[h^\alpha _n, f^\beta _m]=-C^{\alpha
\beta}f^\beta _{m+n}$$   $$[h^\alpha _n, h^\beta _m]={{k}\over {2}}n\delta
^{\alpha \beta}\delta  _{m+n,0}$$

\noindent where the commutators are written down
only for simple roots $\alpha$,  $\beta$.  Semi-infinite homology
$H_{{{\infty}\over 2}+*}(M)$ of this algebra with  coefficients in module
$M$  (provided $k=-2(N+1)$) are defined as the homology of the BRST
complex.  For every generator $g_a$ of the algebra one takes a
ghost-antighost  ($b_a-c^a$)  pair, forming a Clifford algebra {\sl Cliff}:

$$\{b_a,c^d\}=\delta ^d_a.$$

\noindent Picking its Verma module $F_{gh}$ with the
vacuum vector $|0>$, annihilated by  positive modes and $b_{a\ 0}$ and
taking the tensor product (over C)  $F_{gh}\otimes M$ one gets Z-graded
vector space. Then the BRST operator

$$d=\oint (c^a g_a +{1\over 2}
f_{uv}^w c^u c^v b_w)$$

\noindent  introduces the structure of the complex on it.\foot{This
definition resembles that of the usual homologies of Lie algebra,  in
particular, the Cartan--like formula for the boundary operator is the
same in both cases. The only difference is the choice of ``polarization" of
the  corresponding Clifford algebra, i. e. the choice of its highest weight
representation. For the usual homologies we now the definition not
referring  to  any particular ``standard complex", we mean that one given
in terms of the   derived category of $Lie-mod$. It seems very plausible
that there exists the  analogous definition for any choice of the highest
weight representation of   {\sl Cliff} and for semi-infinite homologies in
particular. This fact may be  important to realize because if true, it
sheds light on ``universal" algebraic  nature of BRST for different current
algebras (such as $A^1_N$ and $W_{N-1}$)  ---  it must appear as a derived
functor of Hom (or $\otimes$) in properly  defined derived category of
representations of the algebra.}

Following the lines of motivations of [1,4] we {\it define} the spectrum of
$A^1_N/A^1_N$ coset model as

$$\bigoplus_{\Lambda ,
\Lambda'}H_{{{\infty}\over 2}+*}(L(\Lambda ,k)\otimes W(\Lambda ',
-k-2(N+1))$$

\noindent where $L(\Lambda ,k)$ is the highest weight irreducible
representation of  $A^1_N$  and $W(\Lambda ',-k-2(N+1))$ is a Wakimoto
representation.  It will also be instructive to compute $H_{{{\infty}\over
2}+*}(W(\Lambda  ,k)\otimes W(\Lambda ', -k-2(N+1))$  and
$H_{{{\infty}\over  2}+*}(\tilde{W}(\Lambda ,k)\otimes W(\Lambda ',
-k-2(N+1))$.

To be more precise, let us introduce Wakimoto representation [12,8] $W$ of
$A^1_N$
on the Fock space of the Heisenberg algebra, generated by (the negative modes
of) the currents  $a^i(z), \gamma
^\alpha (z), \beta ^\alpha (z); \ i=1,...,N, \alpha \in \Delta ^+$ ---
positive roots
of $A_N$, with the vacuum vector annihilated by all positive modes {\it and}
all $\beta^\alpha
_0$. The representation is given by means of the bosonisation formulas
(for
the
simple roots):

$$\eqalign{
e^i(z)& = \beta ^{ii+1}-\sum_{j \le i-1}\gamma ^{ji}\beta ^{ji+1} \cr
h^i(z)& = \alpha _+ (a_i,i\partial \phi)+2:\gamma ^{ii+1}\beta
 ^{ii+1}:
-\sum_{j \le i-1}:(\gamma ^{ji}\beta ^{ji}-\gamma ^{ji+1}\beta ^{ji+1}): \cr
  &+\sum_{j \ge i+2}:(\gamma ^{ij}\beta ^{ij}-\gamma ^{i+1j}\beta ^{i+1j}):
\cr
f^i(z)& = - \alpha _+ \gamma ^{ii+1}(a_i,i\partial
\phi)-(k+i-1)\partial
\gamma ^{ii+1} \cr
&-\sum_{j \le i-1}\gamma ^{ji+1}\beta ^{ji}+
\sum_{j \ge i+2}\gamma ^{ij}\beta ^{i+1j}- \cr
  &+
\sum_{j \ge i+2}\gamma ^{ij}\beta ^{i+1j}- :\gamma ^{ii+1}
\bigl(\sum_{j \ge i+1}\gamma ^{ij}\beta ^{ij}-\sum_{j \ge i+2}\gamma
^{i+1j}\beta
^{i+1j}\bigr):
}$$

\noindent where we have used the double script notation $\alpha =ij$ for
the roots of   $A_N$,   taken from the matrix realization of this algebra,
and  $\alpha _+=\sqrt{k+N-1}$.

There is another type of representation we would prefer to call conjugated
Wakimotos.
They can be obtained fixing the Fock modules with a vacuum annihilated by all
positive modes {\it and} all $\gamma ^\alpha _0$, and applying the Chevalley
automorphism
$E^\alpha \leftrightarrow -F^\alpha , H^i \rightarrow -H^i$ to the
bosonisation
above.

Note that the both types of representations are the highest weights. The
other choices for the Fock vacuum give all the other possibilities. All
constructions in this article are shown  for the highest weight
representations and for the convenience of the reader just the final
answers  for   other   types of representations are included.

\subsec{Computation of $H_{{{\infty}\over 2}+*}(\tilde{W}(\Lambda ,k)\otimes
W(\Lambda ', -k-2(N+1))$}

Let's first do the standard trick, extracting
zero modes of Cartan ghosts   $c^i_0,\ b_{i\ 0}\ i=1,\ldots ,N$, to
obtain the so-called relative complex.   Namely, one writes

$$d=\hat{d}+H_{i\ 0} c^i_0+M_i b_{i \ 0}$$   $$H_{i\ 0}=h^M_{i\ 0}+h^T_{i\
0}+\sum_{n \ne 0} f^v_{iu}c^u_{-n}b_{v\ n}$$

$$M_i={{1}\over {2}}
\sum_{n\ne 0} f^i_{uv} c^u_{-n}c^v_n$$

\noindent As $\{d,b^i_0\}=H_{i\ 0}$ and as
the multiplication by $H_0^i$ ( representing  action  of the Cartan
element   of $A^1_N$ on the BRST complex) is, obviously, a morphism of
complexes, there   exists the algebraic   homotopy of the original complex
to its subcomplex $\cap_i {\rm KerH}_{i\ 0}$.       Then it is possible to
restrict ourselves further to the complex

$$ R^\circ =\cap_i ({\rm KerH}_{i\ 0}
\cap {Kerb}_{i\ 0})$$

\noindent which is called the   relative complex.   The ``absolute
homologies", {\sl i.e.,}y that of the original complex, can be restored
by means of the long exact sequence technique.

Now we proceed to computing $H_{{{\infty}\over 2}+*}(\tilde{W}(\Lambda
,k)\otimes   W(\Lambda ', -k-2(N+1))$.            Consider the following
grading on $W\otimes \tilde{W}\otimes F_{gh}$:

$${\rm deg} c^a_n=1 \ {\rm deg} b_{a\ n}=-1 \ {\rm deg} \gamma ^{M\ \alpha}_n=1
\ deg \gamma ^{
T\ \alpha}_n=1$$
$${\rm deg} \beta ^{M\ \alpha}_n=-1 \ {\rm deg} \beta ^{T\ \alpha}_n=-1\  {\rm
deg} a^{i\
-}_n=-1 \
{\rm deg}
a^{+\ i}_n=1$$
where $a^{\pm \ i}_m={1\over {\sqrt{2}}}(a^{M\ i}_m \pm ia^{T\ i}_m)$ and
superscripts ``M" and ``T" distinguish between generators of two bosonic
modules.

As usual, the filtration on the complex $W\otimes \tilde{W}\otimes F_{gh}=
C^\circ$
 is defined by $F^pC^.=\oplus C^i$, where $C^i$ is a subspace of grade $i$,
and so it is also defined on $R^\circ$.
The filtration is compatible with $\hat{d}$: $\hat{d}F^pR^. \subset
F^pR^.$
so one can use it to form a spectral sequence, converging to homologies we
need and
with the first term $E^1_{pq} =H^q(F^pR^./F^{p+1}R^.)$. In other words,
expanding $\hat{d}$ as
$$\hat{d}=\hat{d}_0+\hat{d}_1+\cdots $$
$$\hat{d}_0=\sum_{\alpha \in \Delta ^+}(c^{\alpha}_{-n}\beta^{M\ \alpha}_n+
c^{-\alpha}_{-n}\beta ^{T\ -\alpha}_n)+\sum c^i_{-m}a^{+\ i}_m$$
by the terms of definite degree, one observes that $E^1_{pq}$ is just a
$q$-th homology of operator $\hat{d}_0$, restricted to chains of degree p.

To compute these let us consider an operator
$$ K_0=\sum_{\alpha \in \Delta ^+ }n(b_{\alpha \ n}\gamma ^{M\ -\alpha}_{-
n}+b_{-\alpha \ n}\gamma ^{T\ \alpha}_{-n})+\sum b_{i\ m}a^{+\ i}_{-m}$$
such that $$\{\hat{d}_0,K_0\}=N=
\sum_{\alpha \in \Delta ^+}n(:c^{\alpha}_n b_{\alpha \ -n}:+:\gamma ^{M\ -
\alpha}_n \beta^{M\ \alpha}_{-n}: + :\gamma ^{T\ \alpha}_n \beta^{T\ -
\alpha}_{-n}:)$$
As N is diagonal on our complex, the latter is homotopic to KerN which
is
freely
generated by $\gamma ^{M\ \alpha}_0, \gamma ^{T\ -\alpha}_0, c^{\pm
\alpha}_0$.
Then, a pair of operators $K^+ _\alpha =b_{\alpha \ 0}\gamma ^{M\ -\alpha}_0$,

$K^- _\alpha =b_{-\alpha \ 0}\gamma ^{T\ \alpha}_0$
gives the relations
$$\{\hat{d}_0,K^+_\alpha \} =\beta ^{M\ \alpha}_0 \gamma ^{M -\alpha}_0 +
c^\alpha _0b_{\alpha \ 0}$$
$$\{\hat{d}_0,K^-_\alpha \}=\beta ^{T\ -\alpha}_0 \gamma ^{T \alpha}_0 +
c^{-\alpha}_0 b_{-\alpha \ 0}$$

Again, the right--hand sides of both anticommutators are diagonal on $R^\cdot$
and
define its morphism to itself, and so should annihilate
homologies.
Recalling the definitions of the vacuum vectors of $W$ and $\tilde{W}$ one
immediately sees that the only admissible vector is $\prod_{\alpha \in \Delta
^+}c_0^\alpha |0>$ and that this vector is, indeed, the representative of
$H^n(\hat{d}_0)$, $n={{N(N+1)}\over 2}$.

Thus, only one cell in $E^1_{\cdot \cdot}$ is nontrivial. By the
``dimensional  argument" of the theory of spectral sequences   we conclude,
that this spectral sequence collapses at the first term, {\sl i.e.,}y
$E^\infty _{pq}=E^1_{pq}$, and, at last, $H^i(\hat{d})=\delta _{in}C$.   To
obtain a representative in the only nontrivial class one would generally
apply a ``zig-zag" method, but in this case it is easy to check that
$\prod_{\alpha \in \Delta ^+}c_0^\alpha |0>$   is $\hat{d}$ closed so it is
a true representative already.    Now recall that we are restricted to
$\cap_i {\rm KerH}_{i\ 0}$   which gives: $$\Lambda ^M+\Lambda ^T=\Sigma$$
where $\Sigma$ is a sum   of all positive roots of $A_N$, $\Sigma
=\sum_{\alpha \in \Delta ^+}\alpha   =\sum_i i(N-i+1)\alpha ^i$.

It completes the computation of
$H^{rel}_{{{\infty}\over2}+*}(\tilde{W}(\Lambda   ,k)\otimes W(\Lambda ',
-k-2(N+1))$.

To obtain the ``absolute" homology, one applies (N times) the long exact
sequence.
On performing all intermediate computations, by now pretty standard, we
get the final result (cf. [7,1-3]):
$$H^{abs}_{{{\infty}\over 2}+*}(\tilde{W}(\Lambda ,k)\otimes
W(\Lambda ', -k-2(N+1))=H^{\rm topological}_{i-n}(T^N)$$ where $T^N$ is just a
maximal
torus of $SU(N+1)$. The representatives can be obtained from the monomials
$(c^{i_1}_0 \cdots c^{i_k}_0) \prod_{\alpha \in \Delta ^+}c_0^\alpha |0>$,
$k=0,\ldots ,N$ by the proper twisting.

A few words about other than highest weight representations. Every one
can
be obtained from some HWR by the action  of the Weyl group. For homologies,
the only difference with the computation above will make the choice of the
Fock vacuum. Consider, for example, twisting of $W_M$ by $w \in S_N$. We get a
twisted Wakimoto module $W^w$ whose vacuum
is annihilated by $\beta ^{M\ w\,\alpha}$
if $w\,\alpha \in \Delta ^+$ and by $\beta ^{M\ w\,\alpha}$ in the opposite
case. The same
arguments as above, applied to the new vacuum,
now
single out the vector $\prod_{w\,\alpha \in \Delta ^+}c_0^\alpha |0>$
as a representative of $H^0(\hat{d}_0)$. In this case, however, the full
zig-zag procedure is to be applied to get a true representative in
$H^0(\hat{d})$ ( see also [3] for the case $A_1^1$).

\subsec{Computation of $H_{{{\infty}\over 2}+*}(W(\Lambda ,k)\otimes
W(\Lambda ', -k-2(N+1))$ }

Although the results of the previous section are quite sufficient to
obtain,   via Felder's-like resolution\foot{As the latter procedure appears
to be quite standard {\it in principle}, we   omit any   explicit
construction. Neither do we address an interesting topic of how to find
the   direct representation for the solutions of ``zig-zag" equations}
[9,8],   the homologies of modules like   $L\otimes W$, where $L$ is an
irrep from the main grid of the Kac table, we   also   want to compute the
homologies of the tensor   product of two ``direct" Wakimoto modules. There
are two reasons for our   interest. First is that sometime these would be
useful, and the second, and   more important reason is that, up to now, the
``discrete states" (in a narrow sense)   have been missed. For Virasoro
algebra these states appear (in   $c=1$matter+gravity)   as the nontrivial
homologies of the product of two Fock modules, or, equivalently in
computations of   homologies of modules like $L\otimes W$, where, now, $L$
is an irrep from the  boundary of the Kac table. If one wants to claim that
the spectra of $A^1_N/A^1_N$ coset and $W_N$ gravity are the same thing,
as we do, one should find the counterparts of discrete states in cosets.
This subsection is devoted to this problem.\foot{As in the previous
section, everything will be done explicitly for the HW    representations.}

Let's introduce the new variables
$$ \gamma ^{\pm \ \alpha}_n= {{1}\over {\sqrt{2}}}(\gamma ^{M \ \alpha}_n \pm
\gamma ^{T \ \alpha}_n)$$
$$ \beta ^{\pm \ \alpha}_n= {{1}\over {\sqrt{2}}}(\beta ^{M \ \alpha}_n \pm
\beta ^{T \ \alpha}_n)$$
$$a^{\pm \ i}_n={{1}\over {\sqrt{2}}}(a^{M\ i}_n \pm ia^{T\ i}_n)$$
and assign to them degrees:

$${\rm deg} \beta ^{+\ \alpha}_n=-2 \ {\rm deg} \beta ^{-\ \alpha}_n=-1\  {\rm
deg} a^{+\ i}_n=-2
\ {\rm deg} a^{-\ i}_n=2$$
$${\rm deg} c^a_n=2 \ {\rm deg} b_{a\ n}=-2 \ {\rm deg} \gamma ^{+\ \alpha}_n=2
\ {\rm deg} \gamma ^{
-\ \alpha}_n=1$$
$${\rm deg} a^{\pm \ i}_0=0$$
Following the steps of sect.~1.2, we obtain a spectral sequence with
the first term $E^1_{pq}=H^q(\hat{d}_0)$ --- the $q$-th homology of operator
$\hat{d}_0$\foot{This is homogeneous, degree 0, summand of $\hat{d}$ with
respect to above
defined
grading.},  restricted to subspace of degree $p$.
Now look at $\hat{d}$. It is clear that the only nontrivial contribution to
$\hat{d}_0$ comes from the terms of degree $-2$ in the algebra generators.
The ``purely
ghost" terms like $bcc$ don't contribute to $\hat{d}_0$, either.

Each current $f^a(z)$ is a sum of monomials of the kind
\foot{We deliberately use very sloppy notations,
just to show the structure of the
expression, which is the only important thing.}
$(\prod_i \gamma ^{\alpha _i})\partial \gamma ^\alpha$,
$(\prod_i \gamma ^{\alpha _i})\partial \phi ^i$,
$(\prod_i \gamma ^{\alpha _i})\beta ^\alpha$.
Note that in every term of degree$<1$ in the generators there is an even number
of upper minus signs, the only "odd" terms of degree 1 are $\partial \gamma
^{-\ \alpha},a^{-\ i}\gamma ^{-\ \alpha}$. Using this observation we conclude
that the terms of degree -2 are: $\beta ^{+\ \alpha}_n, a^{+\ i}_n$
Therefore
\eqn\eI{
\hat{d}_0 = \sqrt{2}\big(\sum_{\alpha \in \Delta ^+}c^\alpha _n\beta^{+\
\alpha}_n+
\sum_{n \ne 0} c^i_{-n}a^{-\ i}_n \big)}

There are no terms of degree -1. The terms of degree 0 give rise to the
following form of $\hat{d}_2$:
\eqn\eII{\eqalign{
\hat{d}_2=&[\hat{d}_0,A]+\hat{d'}_2\cr
\hat{d'}_2=&\sum_{a \in \Delta ^+ \cup \{1,\ldots N\}}c^{-\ a}_n
\big(\sum_{\mu +\nu =a}\beta ^{-\ \mu}\gamma ^{-\ \nu}\big)_{-n}\cr
&{1 \over 2}\sum_{\mu ,\nu ,\delta \in \Delta ^+}f_{\mu ,\nu}^\delta (c^{-\
\mu}c^{-\ \nu}b_{- \delta})_0 \cr}}
For example, in the case of $A_2^1$ this operator is:
\eqn\eIII{
\hat{d'}_2=\oint \big(c^{-\ 1}(\beta ^{-\ 2}\gamma ^{-\ 3})+c^{-\ 2}(-\beta
^{-\ 1}\gamma ^{-\ 3})+c^{-\ 1}c^{-\ 2}b_{-\ 3}\big)
}
The operator $\hat{d}_3$ comes from the terms of degree 1 in the generators. It
is
\eqn\eIV{\eqalign{
&\hat{d}_3=-\sum_{\alpha \in \Delta ^+}P_\alpha (n)c^{-\alpha}_{-n}\gamma ^{-\
\alpha}_n \cr
&P_\alpha (n)=2(k+N+1)(n+1)+<\Lambda _M-\Lambda _T,\alpha > \cr
}}
We will prove that the spectral sequence converge at the third term, so we
don't need to know $\hat{d}_4,\hat{d}_5, \ldots$ explicitely (unless we are
interested in the {\it representatives}, but we are not interested in them
here).
The homologies of $\hat{d}_0$ can be computed as in
sec.~2.2.
They are freely generated by (the nonannihillating modes of) $\{c^{-\alpha},
b_{-\alpha},
\gamma ^{-\ \alpha}, \beta ^{-\ \alpha} \}$ and $\{ c^\alpha _0,
\gamma ^{+\ \alpha}_0\}$.
The operator $\hat{d}_2$ acts on the first term of the spectral sequence which
is $H^*_{\hat{d}_0}$. The first summand in \eIV is a commutator with
$\hat{d}_0$. Therefore it acts by zero on $H^*_{\hat{d}_0}$ and can be dropped.

Now look at $\hat{d}_3$. By definition, it only acts on the second term of the
spectral sequence. But from \eIV it follows that it is nilpotent and
(anti)commutes with $\hat{d'}_2$. Morover, $\hat{d'}_2$ is also nilpotent:
\eqn\eV{\eqalign{
& \{\hat{d'}_2,\hat{d'}_2\}=0 \cr
& \{\hat{d}_3,\hat{d}_3\}=0 \cr
& \{\hat{d'}_2,\hat{d}_3\}=0 \cr
}}
 Therefore we may think of the third term of our spectral sequence as of the
cohomology of the bicomplex ($\hat{d}_0$, $\hat{d}_0$, $H^*_{\hat{d}_0}$).
We are free to compute this cohomology by means of the {\it auxillary spectral
sequence}, the first term of which is $H^*_{\hat{d}_3}(H^*_{\hat{d}_0})$.
The latter can easily be computed:

If it never happens that $P_\alpha (r)=0$, the homologies are spanned by zero
modes of all fields, the problem can be
treated as in sec.~2.2 and with the same result.
More interesting is the case when one of $P_\alpha $'s, say, $P_1(r)=0$.
Then $H^*(d_0)$ are generated by both zero modes and $\{c^{-\alpha}_{-r},
b_{-\alpha \ r},
\gamma ^{-\ \alpha}_r, \beta ^{-\ \alpha}_{-r} \}$. Since $H_{i\ 0}$
commutes with $A$, the condition $H_{i\ 0}=0$ can be imposed immediately on
the
elements of $H^*(d_0)$. It gives

$\eqalign{
\Lambda ^i_M+\Lambda ^i_T+\sum_{\alpha \in \Delta ^+} <\alpha, \alpha
_i^\vee>
\big((\#\beta ^{-\ \alpha}-\#\gamma ^{-\
\alpha})+(\#b_{\alpha}-\#c^{\alpha})&\cr
-(\#b_{-\alpha}-\#c^{-\alpha})\big)& =0
}$

The family of operators $K_\alpha =b^{-\alpha}_0\beta^{-\ \alpha}_0$ gives the
relations:
$$\{d_0,K_\alpha \}=P_\alpha (0)(c^{-\alpha}_0 b_{-\alpha \ 0}+\gamma ^{-\
\alpha}_0  \beta ^{-\ \alpha}_0) {=0\; \rm on\; \rm homologies\;}$$
To sum up, there is a restriction
$$\Lambda ^i_M+\Lambda ^i_T +\Sigma +<\alpha, \alpha _i^\vee>
((\#\beta ^{-\ \alpha}_{-r}-\#\gamma ^{-\ \alpha}_r)-
(\#b_{-\alpha \ r}-\#c^{-\alpha}_{-r})=0$$
Solving for both signs of $r$, we obtain

\itemitem{i)} ~

$$\eqalign{\;&{\underline{r>0}}: \; {\rm states\; are\; generated\; by\;}
\beta ^{-\
\alpha}_{-r},c^{-\alpha}_{-r}  \cr
\;&H^0\;: \; {\rm by} \; (\beta ^{-\ \alpha}_{-r})^s|0>\cr
\;&H^1\;: \; {\rm by} \;(\beta ^{-\
\alpha}_{-r})^{s-1}c^{-\alpha}_{-r}|0>\cr}$$

\eqn\eVI{s=-{{\Lambda _M^i+\Lambda _T^i+i(N-i+1)}\over {<\alpha ,\alpha
_i^\vee>}}\ge  0 }

\itemitem{ii)}~

$$\eqalign{\;&{\underline{r<0}} \; {\rm states\; are\; generated\; by\; }
\gamma ^{-\ \alpha}_r,
b_{-\alpha \ r}\cr
\;&H^0 \; : \; {\rm by} \; (\gamma ^{-\ \alpha}_r)^s|0>\cr
\;&H^{-1} \; : \; {\rm by} \;(\gamma ^{-\ \alpha}_r)^{s-1} b_{-\alpha \
r}|0>\cr}$$

\eqn\eVII{s={{\Lambda _M^i+\Lambda _T^i+i(N-i+1)}\over {<\alpha ,\alpha
_i^\vee>}}\ge 0}

Note that \eVI,\eVII should be considered as the equations on $\Lambda
_M+\Lambda _T$ for the fixed s.

It can be then, that $P_\alpha (r)=P_\beta (q)=0$, {\sl etc.,}
(up to N conditions).
For instance, let $P_{\alpha _i}(r_i)=0$, $i=1,\dots,M$, $M \leq N$
and indices
$i$ are so ordered that $r_i \geq 0, \ 1 \leq i \leq M_+$,
$r <0, \ M_++1 \leq i \leq M=M_++M_-$. Then it is easy to convince oneself
that nontrivial homologies arise at the ghost numbers $q,\ \ -M_- \leq q \leq
M_+$, and that $dimH^q={{M!}\over {(M_-+q)!(M-M_- - q)!}}$.

Up to now, we've computed only the first term of the {\it auxillary} spectral
sequence: $H^*_{\hat{d}_3}(H^*_{\hat{d}_0})$.
But we see that all nontrivial elements appear at the same grading;
therefore, by the standard argument, the {\it auxiliary} spectral sequence
converges at {\it the first} term, giving us {\it the third} term of the
{\it main} spectral sequence. Recalling the gradings we see that
$E^3_{pq}=\delta _{p,q}H^q$.
One can draw
$E^3_{..}$ --- see Fig.1
(arrows show the action of $d_3$). We see that the spectral
sequence collapses at the third term --- $E^3_{pq}=E^4_{pq}=\cdots =E^\infty
_{pq}$.
\vskip 3in

At last the relative semi-infinite homology of $A^1_N$ with coefficients in
$W(\Lambda )\otimes W(\Lambda ')$ is:  $$H^{rel}_{{{\infty}\over
{2}}+q}(W(\Lambda )\otimes W(\Lambda '))=  {\bf C}^{D_q}$$
$$D_q={{M!}\over {(M_-+q)!(M-M_- - q)!}} $$   It is clear how to write down
the absolute homologies in this case, see Fig.~2  for an example of the
nontrivial degeneration $P_1(r)=0,\ P_2(s)=0$ for  $A_2^1$. (Each vertex
denotes {\bf C} -- the complex line, the integral number attached to (the
group of) vertices shows the ghost number. To obtain the absolute homology
$H^{abs}_{{{\infty}\over {2}}+q}$ one should take the direct sum of {\bf
C}'s, corresponding to  vertices lying on the $q$-th  plane.)

\newsec{Spectrum of $W_N$ gravity}    \subsec{Some motivations}    In the
previous section the spectrum of $A^1_N/A^1_N$ coset model has been
obtained. If we compare the result for $A^1_1/A^1_1$ coset with that for
${\rm Vir}$-gravity, the correspondence between $A^1_1$- and ${\rm Vir}$-
modules being    given by Hamiltonian reduction, they coincide.

{\bf We suggest that the coincidence of spectra persists for all $A^1_N$ and
$W_{N-1}$.}

There are some points to be clarified.

First of all, it is necessary to define the semi-infinite homology of $W_N$
algebra which is by no means standard as $W_N$ is not a Lie
algebra.\foot{Of course, it is not as difficult to {\it define}
semi-infinite homology    (for instance, see above), as to present a
convenient tool for computing them   --- like BRST complex for Lie
algebra.}

Only recently there has appeared work [10] where claims were made that
the BRST complex does exist for $W_N$ algebra, with ghosts being $b-c$ pairs
of ($b$'s) spins ranging $1,\ldots N$. For $W_3$ algebra the BRST operator is

$$\eqalign{
J(z)&=c^{[2]}(z)\big[\tilde W_L(z) \pm i \tilde W_M(z)\big]
+ c^{[1]}(z)\big[T_L(z)+T_M(z)+T_{gh}^{[1]}(z)+T_{gh}^{[2]}(z)\big]\cr
&+ \big[T_L(z)-T_M(z)\big]b^{[1]}(z)\big(c^{[2]}\del_zc^{[2]}(z)\big)
+ \mu\big(\del_zb^{[1]}(z)\big)c^{[2]}(z)\big(\del^2_zc^{[2]}(z)\big)\cr
&+ \nu b^{[1]}(z)\big(c^{[2]}\del^3_zc^{[2]}c^{[2]}c^{[2]}(z)\big)\ ,
\cr}$$

\noindent where
$\tilde W={1 \over b}W$ and $\mu={3 \over 5}\nu= {1 \over 10{b_L}^2}
(1-17{b_L}^2)$.
Then, it seems to be a problem  to compute explicitly the homologies of
the above-defined
operator, because in this case the bosonisation does not lead to the immediate
answer, as it was for ${\rm Vir}$. (Actually, it can be done for $W_3$, but for
$N>3$ the technical problems accumulate rapidly.)

And after all, suspecting that the semi-infinite homologies of $A^1_N$ and
$W_{N+1}$ coincide, we would seem to be unwise computing them separately
and just comparing the final answers.

In a sense all the problems above can be solved if we understand the role
of Hamiltonian reduction there.

\subsec{Hamiltonian reduction of $A^1_N$ BRST complex}

In [5,6]  Hamiltonian reduction for Wakimoto and irreducible representations
have been defined as the homology $H^0$ of the BRST complex, associated with
the constraints
$$e^\alpha (z)=1\ {\rm if\ \alpha\ is\ a\ simple\ root}$$
$$e^\alpha (z)=0\ {\rm if\ \alpha \ is\ not\ a\ simple\ root}$$

It is important that all these types of
modules are finitely generated over algebra $A^1_N$ (namely, by {\it one}
vector each).

A module like $W_1\otimes W_2$, which is a product of two finitely generated
modules $W_1$ and $W_2$, is infinitely generated, and the definition of
Hamiltonian reduction
for it should be given. It is intuitively obvious (by comparing characters,
for example), that to obtain an expected $F_1\otimes F_2$ --- a product of two
free bosonic representations of $W_{N-1}$ algebra --- one should add twice
as many ghosts as for single $W$ (this must be true of course for any product
of
two finitely generated modules).

It seems natural to make reduction independently on each
factor, {\sl i.e.,}y to take the diagonal ({\sl i.e.,}y $Q=Q_1+Q_2$) cohomology
of the double
complex --- the tensor product of reduction complexes for the factors.

This procedure does work if we only need to reduce a particular module,
without referring to its place in the BRST complex. But we require that the
reduction commute with an $A_N^1$-BRST operator, and it is not difficult to
check that there is no proper modification of $Q_1+Q_2$, commuting with
$Q_{BRST}$.\foot{It is so mainly because in $Q_1+Q_2$ there are not only
symmetric combinations of currents like $J_1^a+J_2^a$ but also the
antisymmetric ones like $J_1^a-J_2^a$.  The latter aren't present in
$Q_{BRST}$ and their commutators with $Q_{BRST}$ cannot  be compensated by
adding ghosts.}

The other problem to deal with is that of the number of ghosts involved. It
seems very natural {\it not} to introduce special reduction ghosts at all,
just   making use of the $A_N^1$-BRST complex ghosts.

Putting all that together, we want to endow the $A_N^1$-BRST complex for
$M_1\otimes M_2$ with
the structure of bicomplex, one differential being $Q_R$, such that $Q_R^2=0$,
and the other one being $Q_{BRST}-Q_R$. The operator $Q_R$ should satisfy
$\{Q_R,Q_{BRST}\}=0$. We require that the complex of $H_{Q_R}^*$ with the
boundary operator $Q_{BRST}-Q_R$ be quasiisomorphic to the $W_N$-BRST complex.

Then we say that $Q_R$ makes the hamiltonian reduction of $A_N^1$-BRST
complex.

Suppose that such an operator exists, and its homology is nontrivial only
in one degree,   say, zero. Then the homology of the bicomplex is
determined by the homology of $W_N$-BRST, on one hand, and is just the
homology of $A_N^1$ by the definition of bicomplex on the other hand. In
this construction, the quasiisomorphism between $A_N^1$- and $W_N$-BRST
complexes is explicit.

Leaving the detailed exposition for the other publication [13], there we
only give an example of such procedure for $\widehat{sl}(2)$ - ${\rm Vir}$
algebras. Even in this simplest case it is already nontrivial. Of course,
everything can be proved just by {\it direct computation of homologies},
but it is exactly what we wish to avoid doing.

Let us consider the operator $$Q_R=\oint c^+(E_1+E_2)+c^0(H_1+H_2+2(b_+c^+
- b_-c^-))$$
 It  satisfies $Q_R^2=0$ and $\{Q_R,Q_{BRST}\}=0$. (But note that it is not a
standard reduction operator!)

The operator $Q_{BRST}-Q_R=\oint c^-(F_1+F_2+c^+b_-)$ acts  on the space
$H^*_{Q_R}$ . We also need to give $H^*_{Q_R}$ the structure of $Vir$-BRST
complex. Look at the ghost sector first. Upon twisting, the conformal
dimension of ghosts $c^-,\ c^0,\ c^+ $ become 1, 0, -1 respectively. $c^-$
has a proper dimension for being $Vir$ ghost, but it is not a singlet with
respect to $sl(2)$.

Bosonisation suggests that in $H^*_{Q_R}$ the ghost $c^-$ gets mixed with
the $\beta ^-, \ \gamma ^-$ pair to give a new scalar with respect to
$sl(2)$ ghost $C$. Instead of doing the bosonisation, we just put $C={c^-
\over {E_1-E_2}}$ and consider an operator  $$ Q_v=\oint :{c^-
(E_1-E_2)^{-1}}( T_1+T_2+T_+ + T_0+{1 \over 2}T_-):$$ where $T_{1,2}$ are
twisted stress energy in matter and Toda sectors, and $T_{\pm ,0}$ are
twisted stress energy for the corresponding ghosts. The total Virasoro
central charge is zero, and it can be checked that $\{Q_V,Q_V\}=0$.
Moreover, $\{Q_V,Q_R\}=0$, so $Q_V$ acts on $H^*_{Q_R}$.

The trickiest part of this definition is the inverse $(E_1-E_2)$. We
define the action of the $\widehat{sl}(2)$ generators on it just by
analytic continuation of the formula for their action on $(E_1-E_2)^n$, for
$n >0$. All the identities with $Q_V$ are the formal consequences of this
definition.\foot{ It is not unreasonable after all that $(E_1-E_2)$ is
invertible on the reduced complex --- we understand that classically we had
$(E_1-E_2)=1$.}

Now let us consider the operator
$$ G={1 \over 2(k+2)}$$
$$ \oint :{c^-(E_1-E_2)^{-1}}
\big( b_+(F_1-F_2)+{1 \over 2} b_0(H_1-H_2) + b_-(E_1-E_2) +
{1 \over 2}\partial b_0 \big):$$
Then there is an identity $$[Q_R, G]=Q_V - {1 \over 2(k+2)}Q_0,$$ therefore on
$H^*_{Q_R}$ holds $Q_V=Q_0$ and, finally
$H^*_{Q_V}(H^*_{Q_R})=H^*_{Q_0}(H^*_{Q_R})$.

The last step is to check that $H^*_{Q_R}$ are nontrivial in only one degree.
But for all representations we discuss in this paper, it is fairly obvious.

\bigbreak\bigskip\bigskip\centerline{{\bf Acknowledgements}}\nobreak

I am grateful to V.~Brazhnikov, V.~Dotsenko, E.~Frenkel, A.~Losev, A.~Mironov,
M.~Olshanetsky
for the encouraging interest to this work, I wish to thank also the organizers
of the Kiev Summer Workshop on 2d gravity, where the main idea has appeared.

I thank M.~Bershadsky  for the explanations of the results of [10] and
very helpful
discussions on the final stage of work.

K.~Pilch red the first version of this paper and pointed out some poorly
explained points and misprints in the computation with two "direct" Wakimoto
modules. It helped me to improve the proof. I thank K.~Pilch very much.

Research supported in part by the Packard Foundation and by NSF grant
PHY-87-14654

\bigbreak\bigskip\bigskip\centerline{{\bf References}}

1. O.~Aharony, O.~Ganor, N.~Sochen, J.~Sonnenschein, S.~Yankielowicz
{\it Physical states in G/G Models and 2d Gravity} TAUP-1961-92.

2. O.~Aharony, J.~Sonnenschein, S.~Yankielowicz
{\it G/G Models and $W_N$ Strings} TAUP-1977-92.~

3. H.~L.~Hu, M.~Yu
{\it  On BRST cohomology of SL(2)/SL(2) gauged WZNW models} AS-ITP-92-32

4. E.~Witten
{\it Comm.~Math.~Phys. {\bf 144}(1992) 189}

5. M.~Bershadsky, H.~Ooguri
{\it Comm.~Math.~Phys. {\bf 126}(1992) 49}

6. B.~Feigin, E.~Frenkel
{\it Comm.~Math.~Phys.~{\bf 128}(1990) 161}

7. B.~H.~Lian, G.~J.~Zuckerman
{\it Phys.~Lett {\bf 254B}(1991) 417,}
{\it Phys.~Lett {\bf 266B}(1991) 21,}
{\it Comm.~Math.~Phys.~{\bf 145}(1992) 54}

P.~Bouwkneght, J.~McCarthy, K.~Pilch
{\it Comm.~Math.~Phys.~{\bf 145}(1992) 541,}
{\it Semi-infinite cohomology in CFT and 2d gravity} CERN-TH.~6646/92

8. P.~Bouwknegt, J.~McCarthy, K.~Pilch
{\it Progr.~Theor.~Phys.~Suppl {\bf 102}(1990) 67}

9. D.~Bernard, G.~Felder
{\it Comm.~Math.~Phys.~{\bf 127}(1990) 145}

10. W.~Lerche, D.~Nemeshansky,M.~Bershadsky, N.~Warner
{\it A BRST Operator for non-critical W-Strings}  HUTP-A034/92

E.~Bershgoeff, A.~Sevrin, X.~Shen
{\it A derivation of the BRST operator for noncritical strings} preprint

11. M.~Thierry-Mieg
{\it Phys.~Lett {\bf B197}(1987) 368}

12. B.~Feigin, E.~Frenkel
{\it Usp.~Mat.~Nauk {\bf 43(5)}(1988) 227}

13. V.~Sadov
{\it The hamiltonian reduction of the BRST complex of
$\widehat{sl}(N)_k/\widehat{sl}(N)_k$-cosets.} HUTP-A006/93

\bye